\documentclass[11pt,letterpaper,notoc]{JHEP3}

\usepackage{url}
\usepackage{graphicx}
\usepackage{latexsym}
\usepackage{amsmath}
\usepackage{amsfonts}
\usepackage{amssymb}
\usepackage{comment}
\def\beq{\begin{eqnarray}}
\def\eeq{\end{eqnarray}}
\def\be{\begin{equation}}
\def\ee{\end{equation}}
\def\bea{\begin{eqnarray}}
\def\eea{\end{eqnarray}}

\def\ltap{\ \raise.3ex\hbox{$<$\kern-.75em\lower1ex\hbox{$\sim$}}\ }
\def\gtap{\ \raise.3ex\hbox{$>$\kern-.75em\lower1ex\hbox{$\sim$}}\ }

\newcommand{\ev}{{\rm eV}}
\newcommand{\kev}{{\rm keV}}
\newcommand{\kevee}{{\rm keVee}}

\newcommand{\gev}{{\rm GeV}}
\newcommand{\tev}{{\rm TeV}}
\newcommand{\mev}{{\rm MeV}}

\title{
Resonant Dark Matter
}

\author{Yang Bai \textnormal{and} Patrick J. Fox\\
Theoretical Physics Department, Fermilab, Batavia, Illinois 60510, USA\\  Email:\email{ bai@fnal.gov, pjfox@fnal.gov}}

\abstract{It is usually assumed that dark matter direct detection is sensitive to a large fraction of the dark matter (DM) velocity distribution.  We propose an alternative form of dark matter-nucleus scattering which only probes a narrow range of DM velocities due to the existence of a resonance, a DM-nucleus bound state, in the scattering - resonant dark matter (rDM).  The scattering cross section becomes highly element dependent, has increased modulation and as a result can explain the DAMA/LIBRA results whilst not being in conflict with other direct detection experiments.  We describe a simple model that realizes the dynamics of rDM, where the DM is the neutral component of a fermionic weak triplet whose charged partners differ in mass by approximately 10 MeV.
 }
\preprint{FERMILAB-PUB-09-415-T}

\begin{document}
\section{Introduction}

Although the fact that a large fraction of the matter in the universe is non-baryonic is beyond doubt, the exact form of the dark matter (DM) and the nature of the dark sector to which it belongs is still shrouded in mystery.  The DM puzzle is under assault simultaneously on several fronts: direct detection experiments search for DM in our galactic halo through its recoil off nuclei in the apparatus, indirect detection experiments look for DM by searching for the standard model (SM) particles produced in DM-DM annihilation in the galactic halo, and collider experiments hope to produce DM in high energy collisions and infer its presence from large amounts of missing transverse energy in events.  We concentrate here on direct detection experiments.

The recent results from DAMA/LIBRA~\cite{Bernabei:2008yi} have clear evidence of an annual modulation in the rate of nuclear recoils in their apparatus.  Taken together with their previous DAMA/NAI results~\cite{Bernabei:2000qi}, they observe this modulation at the $8.2\,\sigma$ confidence level (C.L.).  Furthermore, the peak of this modulation is consistent with June 2$^{\rm nd}$ which is also the peak in the DM velocity relative to the Earth.  The interpretation of the annual modulation as resulting from DM recoiling off nuclei in the DAMA detector is appealing.  However, for the case of a conventional weakly interacting massive particle (WIMP) scattering off nuclei, explaining the DAMA modulation by DM recoils predicts many recoil events at other direct detection experiments, which are not observed.  This apparent tension~\cite{Chang:2008xa,Fairbairn:2008gz,Savage:2008er} between DAMA and other experiments, such as CDMS~\cite{Ahmed:2008eu}, KIMS~\cite{Kim:2008zzn}, XENON~\cite{Angle:2007uj}, CRESST~\cite{Cozzini:2004vd}, and ZEPLIN~\cite{Alner:2007ja,Lebedenko:2008gb}, has lead to new DM explanations where the DM is not a conventional WIMP.

Inelastic dark matter (iDM)~\cite{TuckerSmith:2001hy} proposes that the DM-nucleus interaction is inelastic in nature, the outgoing DM particle is actually an excited state heavier than the DM by $\mathcal{O}(100)\ \kev$.  This changes the kinematics of DM scattering as the DM must have sufficient kinetic energy to be able to up-scatter, which favors the high velocity tail of the DM velocity distribution, and experiments that involve heavy elements.  In particular, for an appropriately chosen splitting, there may be no events in CDMS but there will be events in DAMA and the rest, all of which involve elements at least as heavy as iodine.  The lower cutoff on the DM velocity necessary for an inelastic scatter also means that the amount of modulation in the signal is enhanced, while at the same time the spectrum has a suppression at low recoil energy.  These effects~\cite{Chang:2008gd,MarchRussell:2008dy,Cui:2009xq} help to remove the tension between the DAMA result and the null results of the other experiments, for an analysis that reaches a slightly different conclusion see~\cite{SchmidtHoberg:2009gn}.  In addition, the small splitting in iDM means that the excited state may be very long lived, and the down-scatters that occur when it recoils against a nucleus lead to a novel signal, well outside the usual region of interest that may nonetheless be visible at direct detection experiments~\cite{Finkbeiner:2009mi,Batell:2009vb}.  

A recently proposed alternative to iDM is that the interaction between the DM and the nucleus is momentum dependent~\cite{Feldstein:2009tr,Chang:2009yt}.  Such momentum dependence may come about from a form factor that is present in the coupling of the DM with the SM.  The different DM experiments probe different ranges of momentum transfer and if the momentum dependence is such that outside the range probed by DAMA the form factor is close to zero then DAMA would have increased sensitivity relative to the other experiments.  However, there is still considerable overlap in momentum transfer between the various experiments and so it is not possible, for instance, to arrange for no events in CDMS and only events in DAMA~\cite{Feldstein:2009tr}.  Such a scenario prefers DM of mass around 50 GeV and requires either a non-Maxwellian halo~\cite{Feldstein:2009tr} or spin-dependent couplings~\cite{Chang:2009yt}.

In both of these approaches the scattering cross section is taken to be velocity independent.  For a given recoil energy the scattering takes place for all velocities of DM greater than some lower bound, $v_{min}$, set by the recoil energy; in the case of iDM $v_{min}$ is higher than for elastic scattering due to the mass splitting.  We investigate an alternative possibility, that the DM-nucleus scattering is velocity dependent, and in particular that the scattering takes place through the production of a resonance, the scattering cross section then has a Breit-Wigner form.  The resonant form of the scattering, resonant dark matter (rDM), means that for given DM and target masses only a narrow range of velocities, around the resonance velocity, will actually scatter and leave a signal in the detectors.  Since only a small portion of the whole DM velocity distribution is probed there is increased modulation of the recoil spectrum, despite the scattering being elastic in nature.  As emphasized in~\cite{Chang:2008gd} this helps to reconcile the DAMA results with the null experiments.   In the concrete realization of rDM that we present here the resonance is a bound state of a nucleus and a charged partner of the DM, and the mass of the bound state is very sensitive to the particular target nuclei in the direct detection experiment.  The velocity of DM in the Earth's frame is restricted to lie in the range $0~\mathrm{km/s} \ltap v_{DM}\ltap 800$~km/s.  If the corresponding resonance velocity does not lie below $\sim 800$ km/s there will be no resonant scattering, and the signal of DM recoils will be greatly suppressed.  This high element sensitivity means that, unlike the two approaches outlined above, it is possible in rDM that iodine is the \emph{only} element for which direct detection could occur.

Before focusing on the particular form of the resonance as a DM-nucleus bound state, we first describe rDM in a model independent fashion, emphasizing the basic features should a resonance exist, in Section~\ref{sec:ResonanceEffects}.  We go on to show how the two experiments involving iodine, DAMA and KIMS, can be explained simultaneously and discuss in general the allowed parameter space.  Then in Section~\ref{sec:model}, we introduce an explicit model, based on the work of Pospelov and Ritz~\cite{Pospelov:2008qx}, that realizes many of the necessary features.  We demonstrate the sensitivity to the target element and discuss the various possible signatures and relevant constraints in Section~\ref{sec:elemental}.  Finally, we conclude in Section~\ref{sec:conclusion}.

\section{Resonance Effects}
\label{sec:ResonanceEffects}

In this section, we consider other possible effects, which can dramatically change the results from traditional DM-nucleus elastic scattering cross section calculations.  The detection rate per unit detector mass at a DM direct detection experiment is given by~\cite{Lewin:1995rx}
\be
\label{eq:diffrate}
\frac{dR}{dE_R}=\frac{N_T\,m_N\,\rho_\chi}{2\,\mu_{N\chi}^2\,m_\chi} \int_{v_{min}} d^3\vec{v} \,\frac{f(\vec{v},\vec{v}_E)}{v}\,\sigma_N\,F^2(E_R)~,
\ee
where $m_N\approx A\,m_P$ is the nucleus mass with $m_P$ the proton mass and $A$ the atomic number; $F(E_R)$ is the nuclear form factor and accounts for the fact that the cross section drops as one moves away from zero momentum transfer; the two-parameter Fermi charge distribution is used to calculate $F(E_R)$ throughout this paper~\cite{Duda:2006uk}; $N_T$ is the number of target nuclei per unit mass, given by $N_T=N_A/A$ with Avogadro's number, $N_A=6.02\times10^{26}$~kg$^{-1}$; $\sigma_N$ is the cross section to scatter of a nucleus, and $\mu_{N\chi}$ is the reduced mass of the DM-nucleus system.  The DM mass is $m_\chi$ and we take the local DM density to be $\rho_\chi = 0.3$~GeV/cm$^3$. The velocity of the dark matter onto the (Earth-borne) target is $\vec{v}$.  The Earth's velocity in the galactic frame, $\vec{v}_E$, is the sum of the Earth's motion around the Sun \cite{Lewin:1995rx} and the Sun's motion in the galaxy \cite{Dehnen:1997cq}.  We assume the WIMP velocity distribution is Maxwell-Boltzmann with velocity dispersion $v_0=220$~km/s.  Thus,
\be
f(\vec{v}, \vec{v}_E)=\frac{1}{(\pi\,v^2_0)^{3/2}}\,e^{-(\vec{v}+\vec{v}_E)^2/v^2_0}~.
\ee
As a function of time in the galactic frame, the Earth's velocity is  $v_E\approx 227 + 14.4\cos{[2\pi\left(\frac{t-t_0}{T}\right)]}$ ~km/s, with $T=1$~year and $t_0$ is around June 2$^{\rm nd}$. The DM velocity distribution is cut-off at the galactic escape velocity.  Thus, the upper limit of the integration in (\ref{eq:diffrate}) is given by $|\vec{v}+\vec{v}_E| \le v_{esc}$, and the lower limit, since we will consider elastic scatters, is given by
\be
v_{min}=\sqrt{\frac{m_N E_R}{2\,\mu_{N\chi}^2}}~.
\ee
The current allowed range for the galactic escape velocity \cite{Smith:2006ym} is $498$ km/s $\le v_{esc}\le$ 608 km/s.  For concreteness we set $v_{esc}=500$ km/s.  Increasing this value slightly increases our allowed parameter space, but the general features remain unchanged.  
Because of different energy detection efficiencies for different detectors, a quench factor $f_q$ is introduced to relate the observed recoil energy, $\bar{E}_R$, to the actual recoil energy $E_R$, $E_R=\bar{E}_R/f_q$.  This allows one to convert Eq.~(\ref{eq:diffrate}) to the experimental differential spectrums as $dR/d\bar{E}_R = 1/f_q\,dR/dE_R$. For example, we take the quench factor $f_q=0.085$  for the iodine element in the DAMA experiment.
\FIGURE[t!]{
\includegraphics[width=0.6\textwidth]
{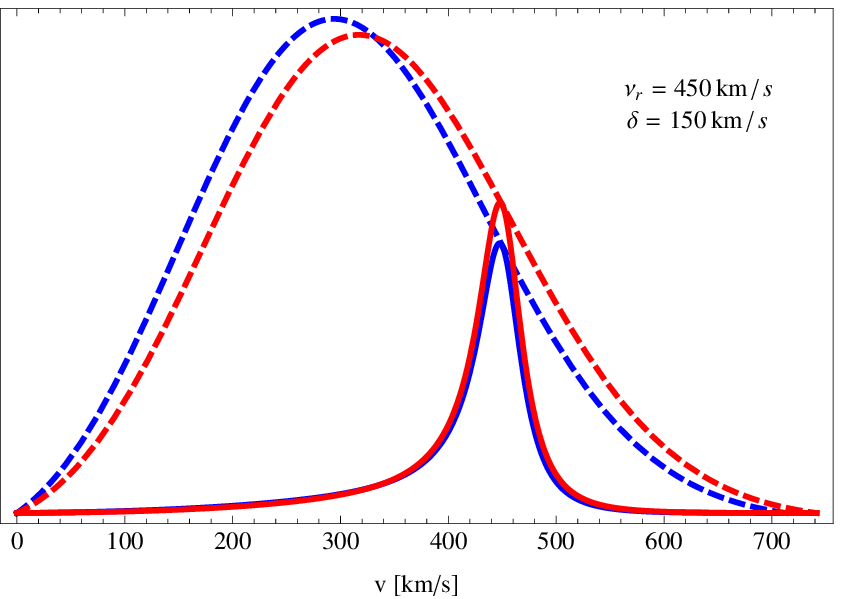} 
\caption{DM velocity distribution after angular integration in the summer (red) and winter (blue) for the usual Maxwell-Boltzmann distribution (dashed) and for the case with a resonance (solid) at 450 km/s with width 150 km/s, the escape velocity was taken to be 500 km/s.}
\label{fig:velocitydistribution}}

In the usual calculation the nuclear cross section $\sigma_N$ is related to the nucleon scattering cross section, $\sigma_p$, by,
\be
\sigma_N=\frac{(Z f_p+(A-Z)f_n)^2}{f_p^2}  \frac{\mu^2_{N\chi}}{\mu^2_{n\chi}} \sigma_p~,
\ee 
where $f_{p,n}$ are the coupling strengths of DM to protons and neutrons and $\mu_{n\chi}$ is the DM-nucleon reduced mass.  Here however, we wish to work explicitly with the nuclear scattering cross section, and leave relating it to the microscopic Lagrangian to later, section \ref{sec:model}.  In particular, in the usual approach $\sigma_p$ is velocity, and element independent.  We will see that in rDM these statements are no longer true.

In rDM, the DM or its gauge partner, forms a short-lived bound state with the target nucleus. The mass of the bound state is denoted as $m_r$.  In this case the DM-nucleus \emph{elastic} scattering cross section has a resonant structure.  In the non-relativistic limit, one has $s=(m_\chi+m_N)^2+m_\chi m_N v^2$. For $\sqrt{s}$ close to the resonance mass, a familiar formula is obtained,
\be
\sigma_N=\frac{2J_r+1}{(2s_\chi+1)(2s_N+1)}\,\frac{\pi}{k^2}\,\frac{\Gamma^2_{r\rightarrow \chi N}}{(E-m_r)^2+\Gamma_{tot}^2/4}~,
\label{eq:resonancexsec}
\ee
where $E=\sqrt{s}$  is the center of mass energy; $s_\chi$ and $s_N$ are the spins of the dark matter and the target nucleus; $J_r$ is the total angular momentum of the resonant bound state. In the non-relativistic limit, the scattering process is dominated by the $s$-wave, so a selection rule, $\overrightarrow{J_r}= \overrightarrow{s_\chi} +\overrightarrow{s_N}$, applies to the accessible bound state.  
$\Gamma_{r\rightarrow \chi N}$ is the partial width of the boundstate decaying into $\chi$ plus $N$ and is a function of the centre of mass energy.  The total width $\Gamma_{tot}$ may be larger than this width due to the existence of other decay modes, we will discuss this in more detail in Section~\ref{sec:model}.  The momentum of the DM in the center of momentum frame is $k=\mu_{N\chi}\, v$.
Note that if there exists more than one resonance, the cross section is the sum over all resonances, each given by (\ref{eq:resonancexsec}). 
Since the DM is non-relativistic, we can rewrite the cross section as a resonance in velocity, 
\be
\sigma_N =\sigma_0\ \frac{v_r^2}{v^2}\,\frac{\delta^2/\pi}{(v^2-v_r^2)^2+\delta^4}\,.
\label{eq:xsecformula}
\ee
Here, the normalization is 
\be
\sigma_0= \frac{2J_r+1}{(2s_\chi+1)(2s_N+1)}\,\frac{4\pi^2}{\mu_{\chi N}^2} 
\frac{\delta^2}{v_r^2}\frac{\Gamma^2_{r\rightarrow \chi N}}{\Gamma_{tot}^2}, 
\ee
and the resonance velocity and width are given by,
\bea
v_r^2=\frac{2\,(m_r-m_\chi-m_N)}{\mu_{\chi N}}\,,\qquad \delta^4=\frac{\Gamma_{tot}^2}{\mu^2_{\chi N}}~.
\label{eq:vrdelta}
\eea
For a narrow resonance the widths are constant, but for a wide resonance one must take into account their dependence on the velocity. In the narrow resonance case with $\delta \ll v_r$, Eq.~(\ref{eq:xsecformula}) is well approximated by a delta function and is 
\be
\sigma_N =\sigma_0\, \delta(v^2-v_r^2)\,.
\label{eq:xsecformula2}
\ee
\FIGURE[t!]{
\includegraphics[width=0.6\textwidth]
{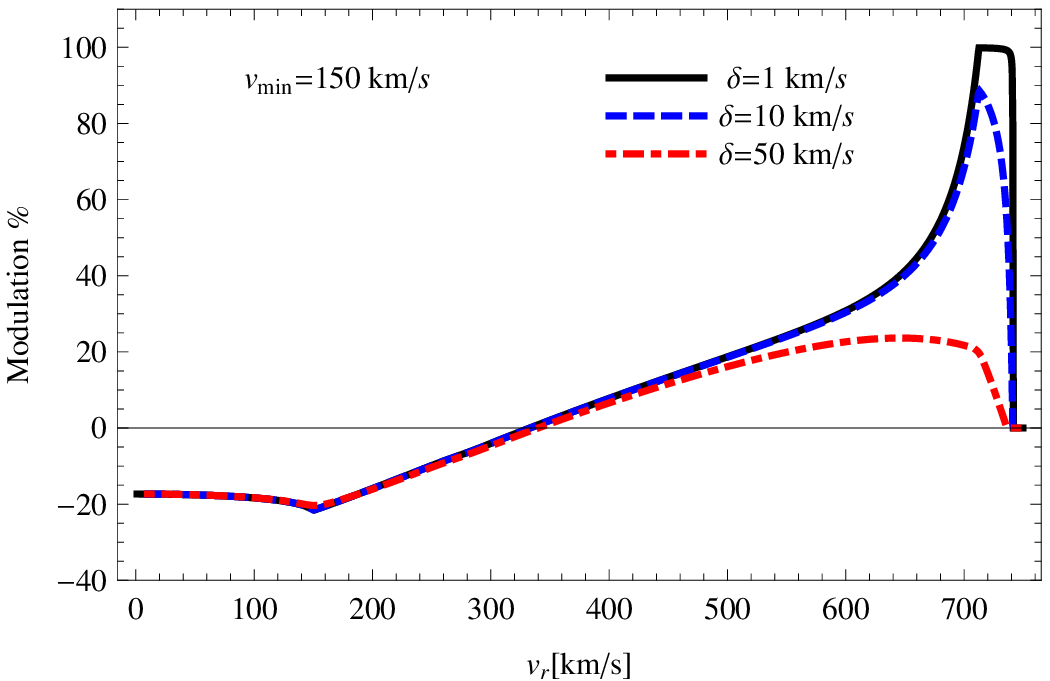} 
\caption{The ratio of modulation as a function of the resonance velocity $v_r$. We take this ratio as the summer event rate minus the winter rate relative to the summer rate. The $v_{min}$ is chosen to be 150~km/s and $v_{esc}=500$ km/s. The black (solid), blue (dashed) and red (dot-dashed) lines are for widths of $\delta= 1, 10, 50$~km/s, respectively.}
\label{fig:modulationratio}}

For the existence of a resonance to have an observable effect the resonant velocity, $v_r$, must be low enough that some DM particles have this velocity and the resonance must be narrower than the range of the DM velocity distribution, $\delta \ltap v_{esc}$.  This second condition means that the resonance is relatively narrow, $\Gamma_{tot}\ltap 1\ \mev$.  If these two conditions hold then the presence of the resonance effectively picks out only part of the Maxwell-Boltzmann distribution, Fig.~\ref{fig:velocitydistribution}. In doing so the resonance increases the amount of modulation in the signal, since the signal now comes from a narrow range of velocities over which the summer and winter rates are discrepant, and there is no averaging over the whole velocity distribution.  

The total counting rate can be separated into two parts $S = S_0 + S_m \cos{[2\pi(t-t_0)/T]}$, with $S_0$ as the unmodulated rate and $S_m$ as the modulated rate. The amount of modulation, and the sign, now depend very sensitively on the values of $v_r$ and $\delta$.  This feature of increased modulation is similar to inelastic DM \cite{TuckerSmith:2001hy,Chang:2008gd}, which also happens due to a restriction on which part of the Maxwell-Boltzmann distribution is probed, but occurs here despite the scattering being elastic in nature.  

In Fig.~\ref{fig:modulationratio} we show a plot of the fraction of modulation, defined as the ratio of the difference of the summer and winter rates to the summer rate, as a function of the resonance velocity for three choices of the resonance width.  For illustration purposes we take the DM mass to be 500 GeV and we consider the ratio in the bin that corresponds to $3~\kevee < \bar{E}_R < 3.5~\kevee$ at DAMA.  This is representative of what happens in other bins, and for other DM masses.  For this bin, the corresponding $v_{\rm min}$ is $150$~km/s.
  
From Fig.~\ref{fig:modulationratio}, we see that with sufficiently small $\delta$ the ratio can be as large as 100\%.  If the resonance velocity is below the peak of the Maxwell-Boltzmann distribution it is possible to have a modulation ratio that is negative (meaning the winter time has more data than summer time) since below the peak the distribution is higher in the winter than in the summer.
\FIGURE[t!]{
\centerline{
\includegraphics[width=0.48\textwidth]{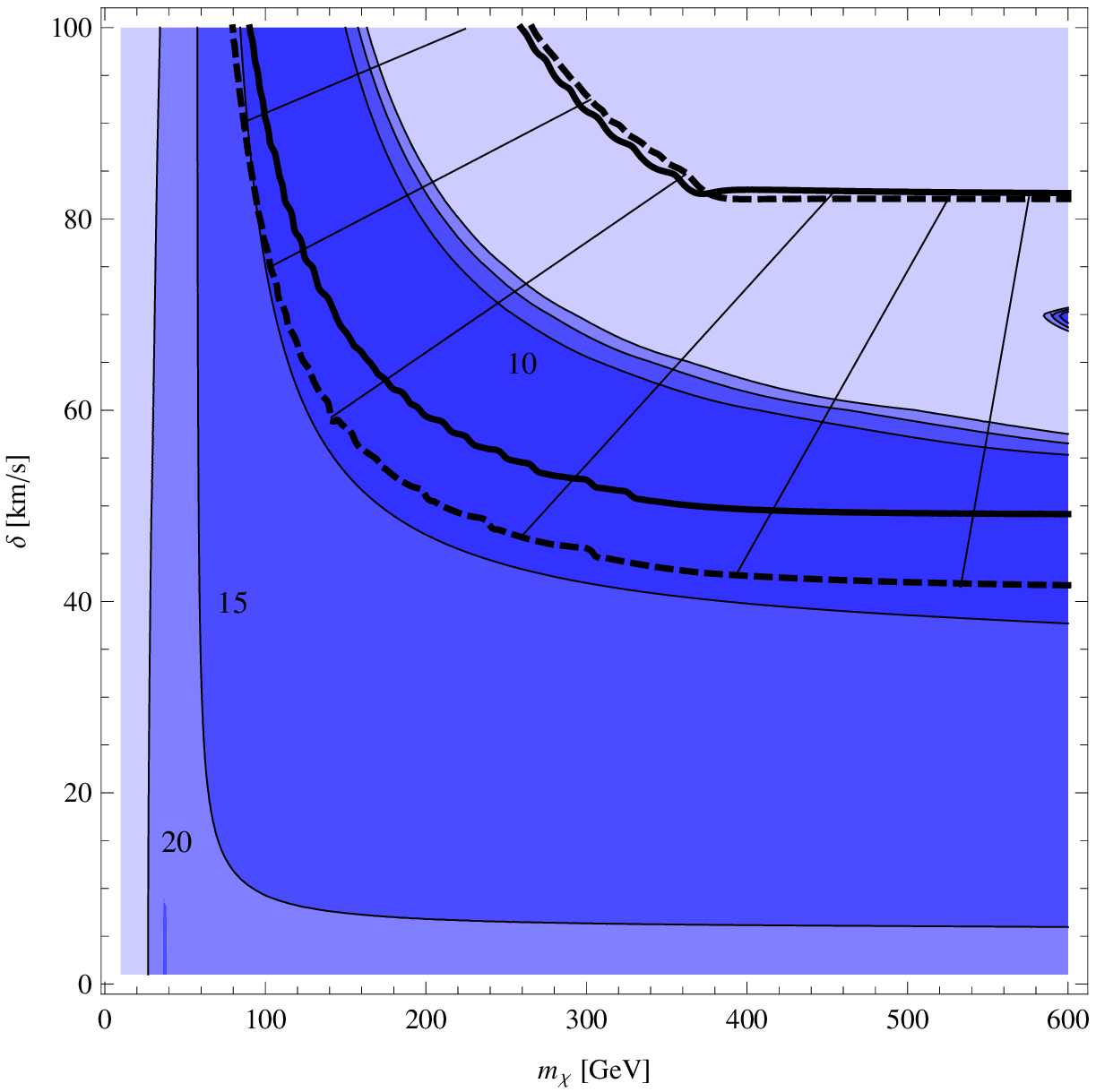} \hspace{3mm}
\includegraphics[width=0.48\textwidth]
{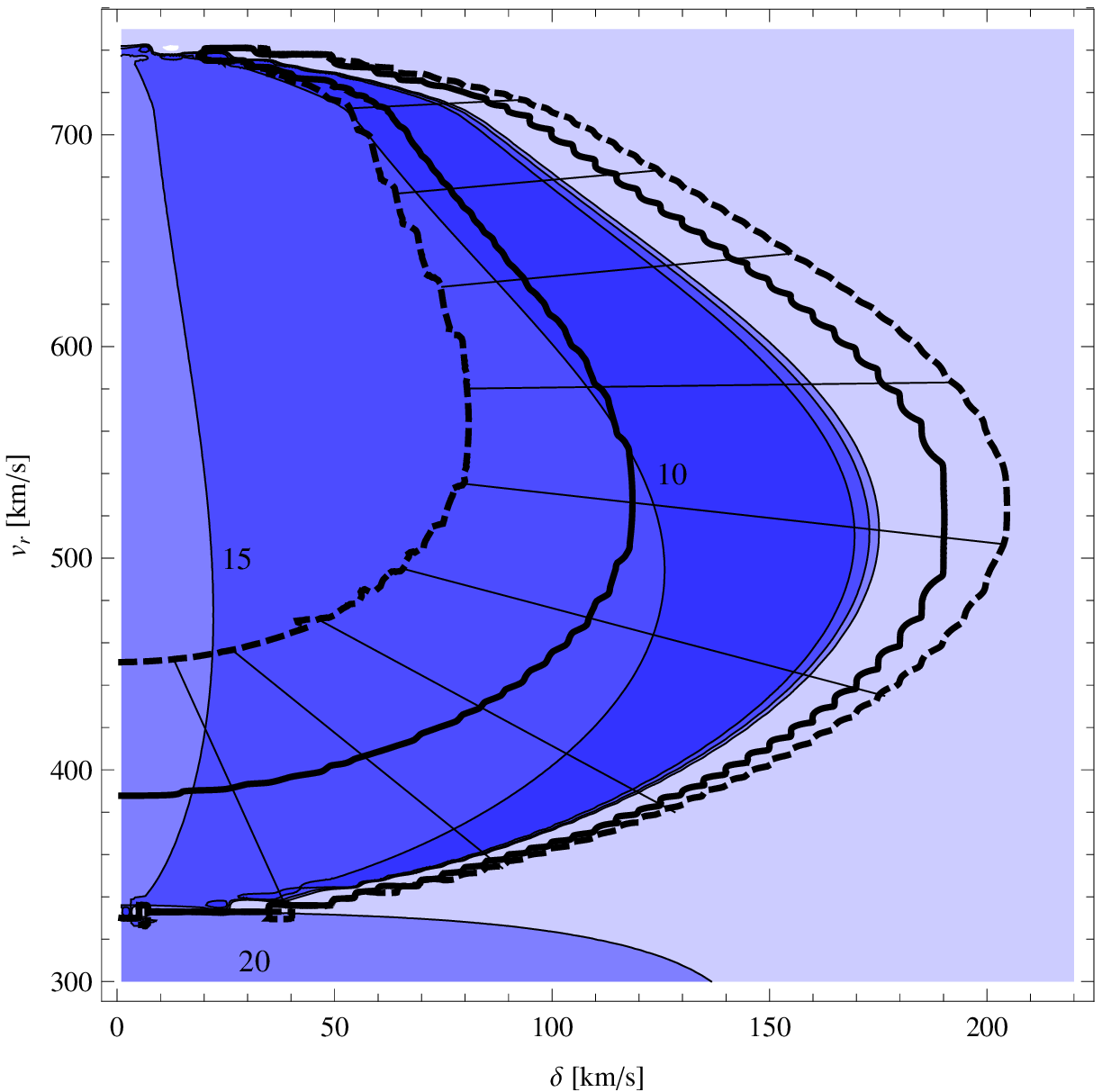} 
}
\caption{Allowed regions of parameter space.  The left plot has $v_r=725$ km/s and the right plot has $m_\chi=500\ \gev$.  In both cases the shaded contours denote the value of $\chi^2$ for a fit to the first 12 bins of the DAMA modulated data, and the region between the solid (dashed) black lines exceed the DAMA unmodulated (KIMS) data at 90\% C.L.}
\label{fig:allowedregions}}

One candidate for the resonance is a DM-nucleus bound state, we will describe a model with just such an object in Section~\ref{sec:model}, in which case the resonant velocity may be highly element dependent.  This opens up the possibility that iodine maybe the only element with an open resonance, and in other experiments the signal rate would be suppressed by $\sim(\delta/v_r)^4\ltap 10^{-4}$. In this case only the DAMA and KIMS experiments, which both contain iodine, would be expected to observe a signal.  We will show that because of the modulation enhancement in rDM it is possible to simultaneously explain the modulated DAMA results without contradicting either the DAMA unmodulated results or the KIMS results.
\FIGURE[t!]{
\includegraphics[width=0.47\textwidth]{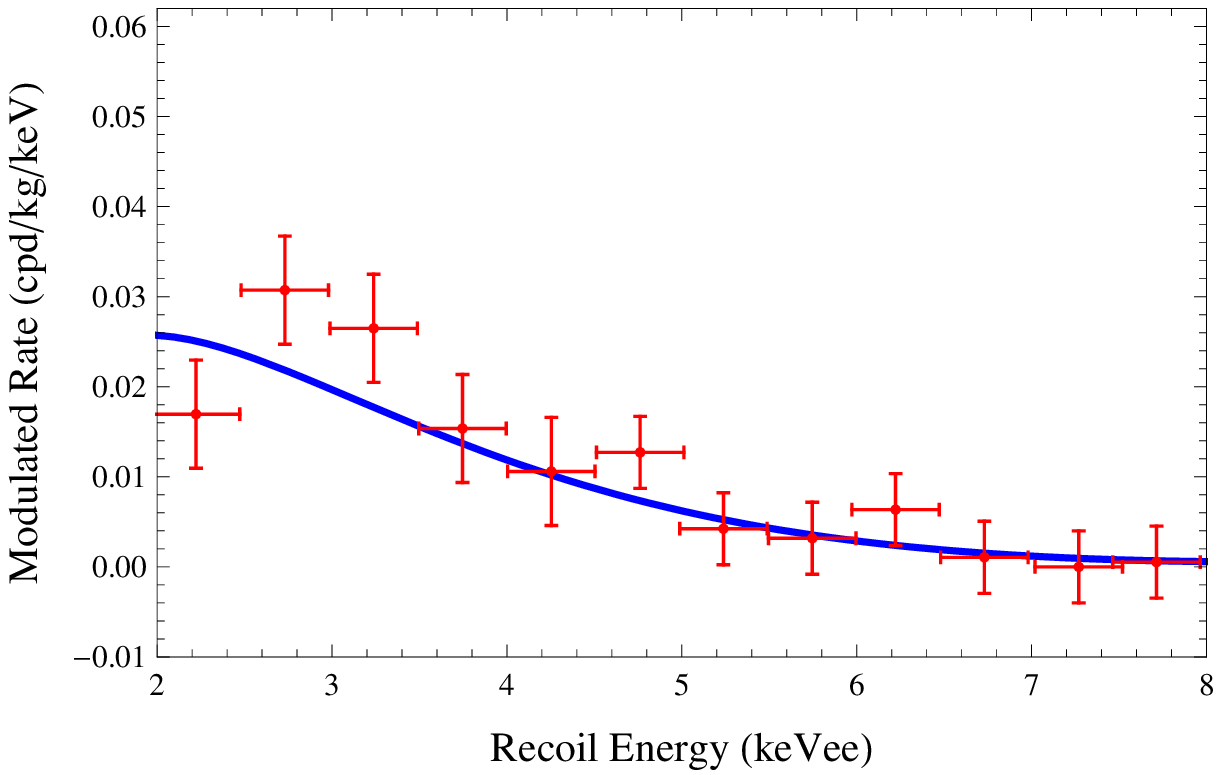}  \hspace{3mm}
\includegraphics[width=0.45\textwidth]
{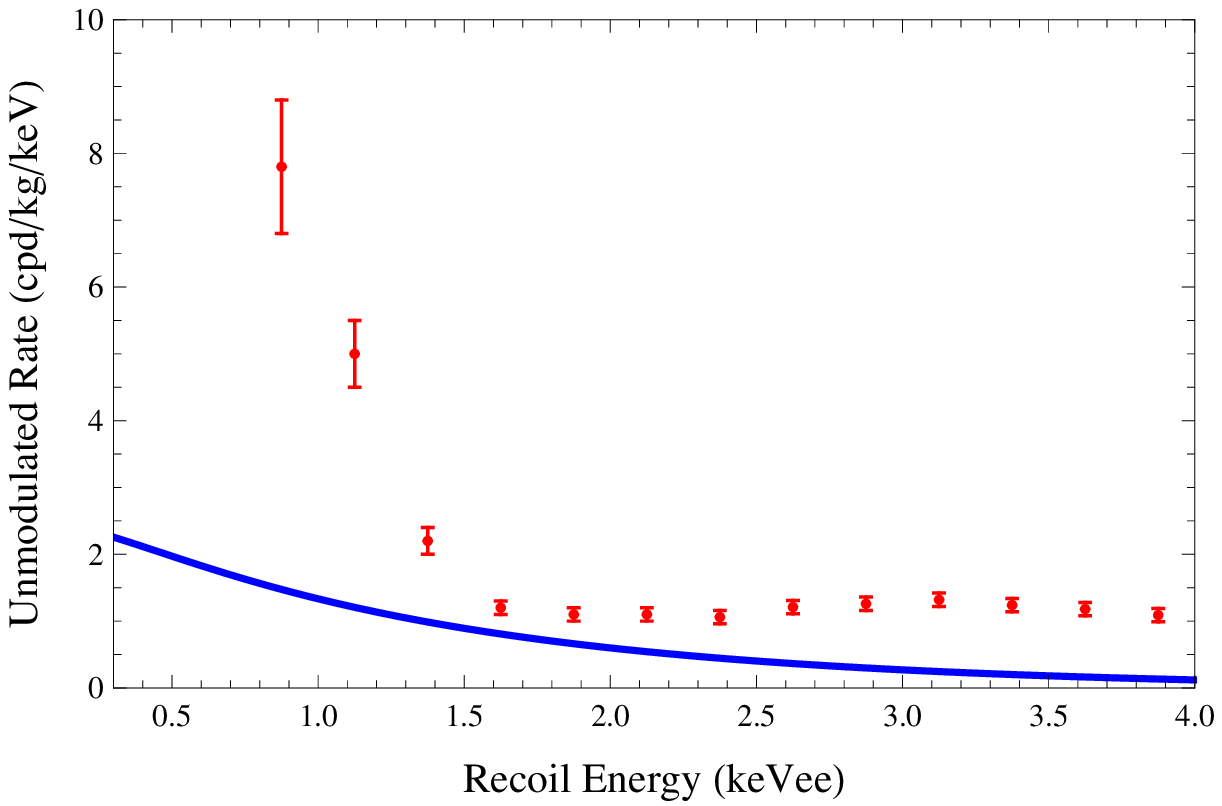} 
\caption{Left panel: Model predictions from a single resonance and the DAMA modulated date. Right panel: Same as the left panel, but for the unmodulated rate. For both plots, we have used $m_\chi=500$~GeV, $\sigma_0 = 1.34$~pb, $v_r=725$~km/s and $\delta=40$~km/s. This model point is also allowed by the KIMS experiment.}\label{fig:Goodone}}

Independent of the specific details of a model, we have four parameters to describe the modulation spectrum~\footnote{In the concrete model described in Sec.~\ref{sec:model}, some of these parameters are related and  there are only two independent parameters in total.}: $m_\chi$, $v_r$, $\delta$ and $\sigma_0$. Because of the uncertainty in the number of parameters in a realistic model, we present the total $\chi^2$ to describe the goodness of fit for the first 12 bins, $2\,\mathrm{keVee} \le \bar{E}_R \le 8$ keVee, of the DAMA modulated spectrum. We also consider constraints from the ``single-hit" unmodulated spectrum in DAMA and the results from KIMS experiment. The constraints from other experiments are model-dependent and will be discussed in Section~\ref{sec:model}.

The allowed parameter space is shown in Fig.~\ref{fig:allowedregions},
where we treat the overall cross section $\sigma_0$ as a floating parameter to minimize the $\chi^2$ for a fit to the DAMA modulated spectrum, for a given set of $m_\chi$, $v_r$ and $\delta$. The $90\%$ C.L. exclusion region, enclosed by the black solid and the black dashed lines, are from the unmodulated data of DAMA and KIMS, respectively.  The first 22 bins (up to 8~MeV) of DAMA and all 8 bins of KIMS are included in this analysis. A constant quench factor, 0.085, for Iodine is used for the DAMA experiment. For KIMS,  we interpolate the energy dependent quench factor~\cite{Park:2002jr} as $f_q(E_R)=0.1\,e^{-0.0135\,E_R}+0.06$ for $E_R$ in~keV, or equivalently, $f_q(\bar{E}_R)=e^{-(\bar{E}_R^2+5 \bar{E}_R)/90}+0.06$ for $\bar{E}_R$ in~keVee. We take the statistically averaged values of the four crystals in KIMS as the experimental input. 

From the left panel of Fig.~\ref{fig:allowedregions}, we can see that the fit to DAMA modulated data is almost independent of the DM mass, once $m_\chi$ is above 300 GeV. The KIMS constraint is stronger than the DAMA unmodulated one. Together, the parameter space with a total $\chi^2$ below $10$ is almost ruled out, but a large area of parameter space with $\chi^2$ between 10 and 15 is allowed. One can also see that a larger $v_r$ and a smaller $\delta$ can more easily evade the constraints, which is due to the enhanced modulation effects for this part of parameter space.  For illustration purpose, we show one allowed point of the parameter space in Fig.~\ref{fig:Goodone}, which has $\chi^2=9.5$, and another in Fig.~\ref{fig:single}, which has $\chi^2=15.2$~. For Fig.~\ref{fig:allowedregions}-\ref{fig:single} we have used $v_{esc}=500$~km/s and, for DAMA, have taken into account the detector energy resolution, by smearing with a Gaussian distribution with $\sigma(E)/E=0.448/\sqrt{E}+0.0091$~\cite{Bernabei:2008yh} with $E$ in keVee.

\FIGURE[t!]{
\includegraphics[width=0.465\textwidth]{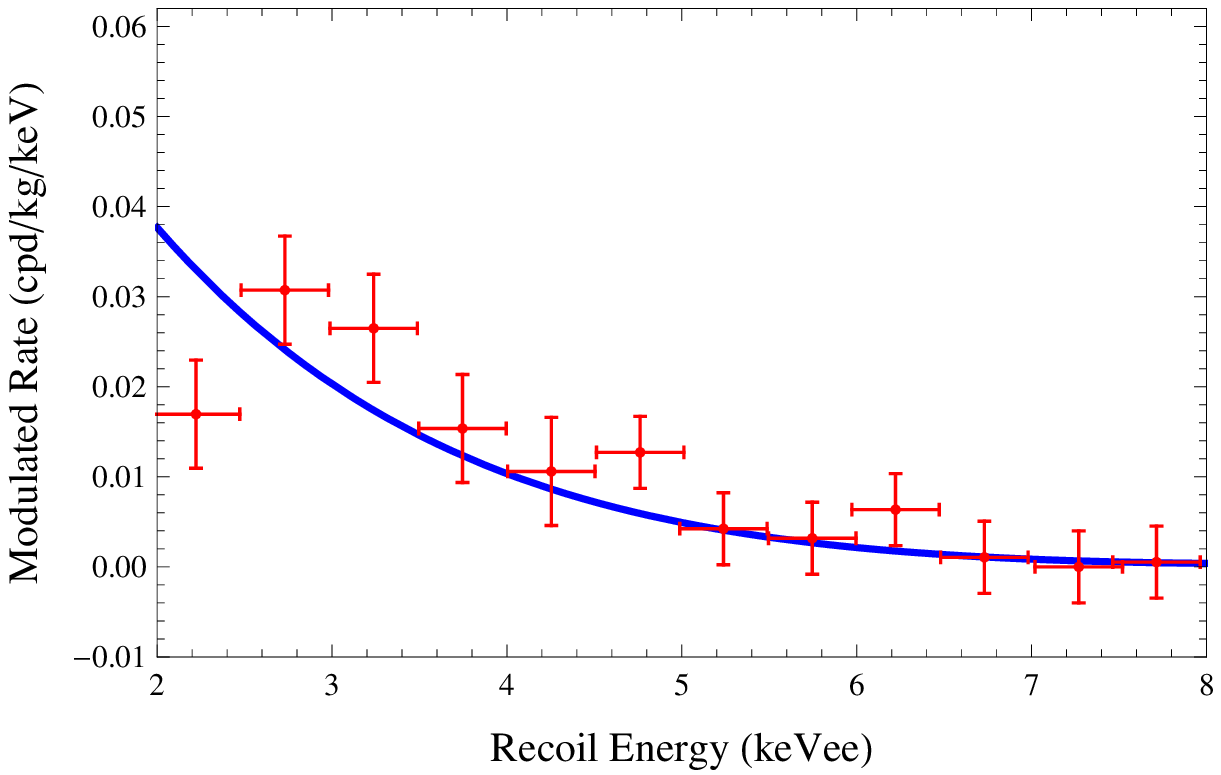} \hspace{3mm}
\includegraphics[width=0.45\textwidth]{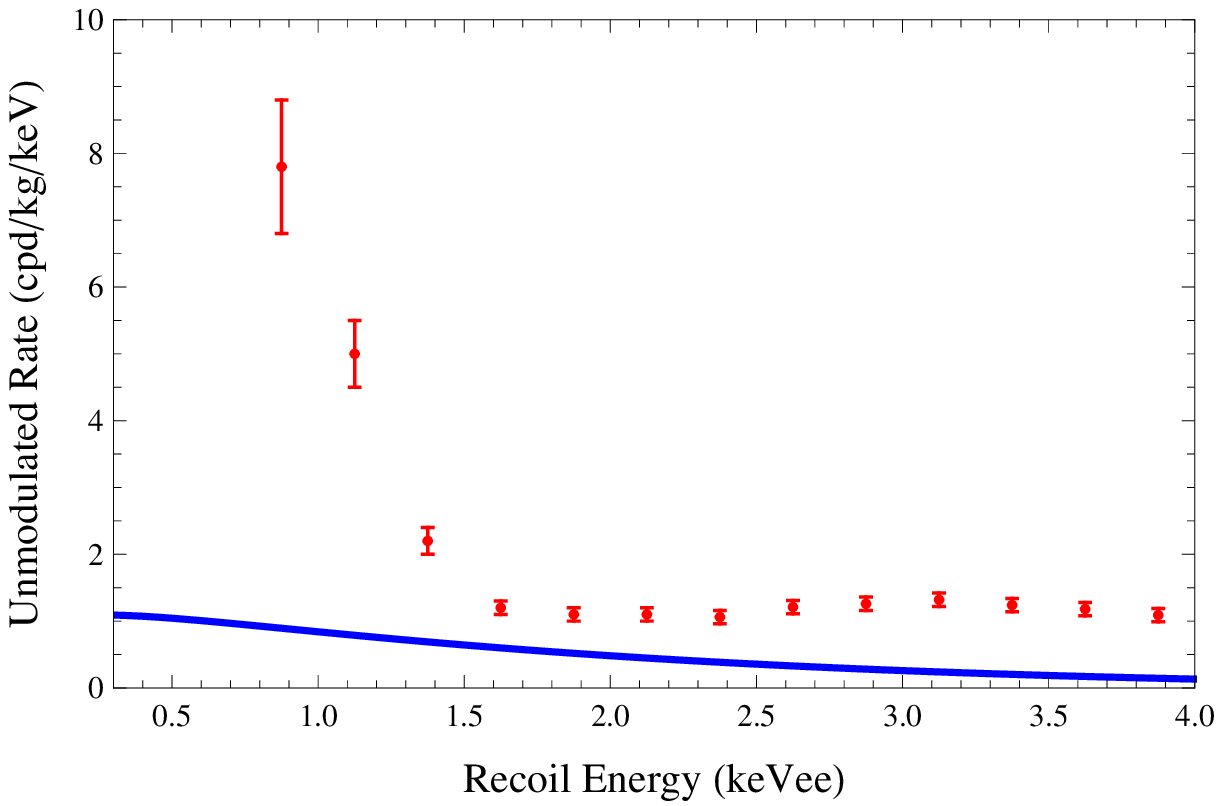} 
\caption{The same as Fig.~\ref{fig:Goodone}, but for $m_\chi=500$~GeV, $\sigma_0 = 0.025$~pb, $v_r=470$~km/s and $\delta=0.075$~km/s.}
\label{fig:single}}
%

\section{A Model}
\label{sec:model}

As a realization of rDM we consider the case of a fermionic WIMP, $\chi^0$, which is nearly degenerate in mass with a charged partner, $\chi^\pm$, ~\cite{Pospelov:2008qx}, we will be interested in splittings of order $10-100$ MeV.  We also introduce a ${\cal Z}_2$ symmetry under which the dark matter particle is odd and thus $\chi^0$, the lightest parity-odd particle, is a stable particle.  In order to suppress the spin-independent coupling of $\chi^0$ to nuclei through Z boson exchange we take $\chi$ to transform as $(3, 0)$ under $SU(2)_W\times U(1)_Y$.  

The splitting of the charged from the neutral component can come from two sources.  There may be higher dimension operators such as $(\overline{\chi}T^a\chi)(H^\dagger T^a H)$ suppressed by some scale $\Lambda$, which contribute $v^2/\Lambda\equiv \Delta_{UV}$.  There are also loop generated contributions. After electroweak symmetry breaking the charged components of $\chi$ are split from the neutral component by electromagnetic radiative corrections, the size of this splitting~\cite{Cirelli:2005uq,Essig:2007az} is independent of the dark matter mass for $m_\chi \gg M_W$ and is around $\alpha_2 M_W \sin^2\frac{\theta_W}{2}\approx 166$~MeV.  The higher dimension operator is comparable to the radiative correction for $\Lambda\sim 10^6 -10^7$~GeV. For latter analysis, we require a mild cancellation between those two contributions and have
\be
m_{\chi^\pm}-m_\chi \equiv \Delta =166\ \mev - \Delta_{\rm UV} \sim \mathcal{O}(10) ~\mev~.
\ee
\FIGURE[htb!]{
\includegraphics[width=0.6\textwidth]{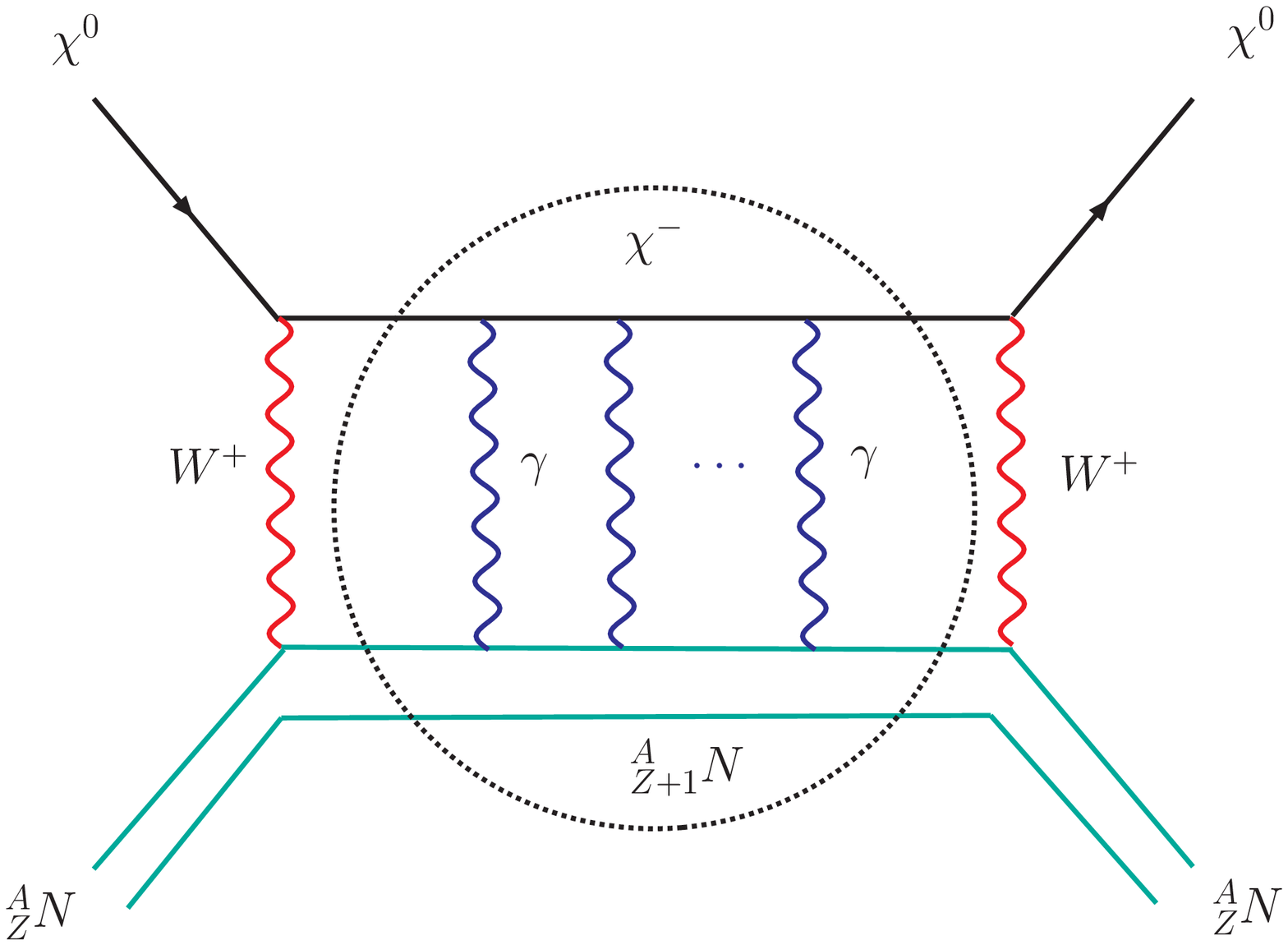} 
\caption{The Feynman diagram of the DM elastic scattering off a nucleus $^A_ZN$ by exchanging a boundstate of ($\chi^-$, $_{Z+1}^AN$).}
\label{fig:feynman}}
The small splitting between $\chi^\pm$ and $\chi^0$ results in a long lifetime for the charged state.  The width for the decay $\chi^\pm\rightarrow \chi^0 e^\pm \nu$ is
\be
\Gamma \approx \frac{1}{15\,\pi^3} G_F^2\Delta^5 \approx \left(\frac{\Delta}{15~\mev} \right)^5\times 2\times 10^{-13}\ \ev~,
\label{eq:decaywidth}
\ee
neglecting the electron mass. The lifetime of $\chi^\pm$ is $\tau\approx 3\times 10^{-3}$~s for $\Delta = 15$~MeV. Because of the almost degenerate masses of $\chi^\pm$ and $\chi^0$, they have approximately the same amount of thermal relic abundance. The charged particles decay to $\chi^0$ shortly after the freeze-out time. To satisfy the observed value of dark matter relic abundance, the dark matter mass should be around 2~TeV~\cite{Cirelli:2005uq}. Keeping in mind that the dark matter abundance may also originate from other non-thermal processes, we will treat the dark matter mass as a free parameter in this paper. We also note that the lifetime of $\chi^\pm$ is far less than one second. Therefore, Big Bang nucleosynthesis is unaffected in the model.

The lifetime of $\chi^\pm$ is fairly long from the collider point of view. If produced in a collider, $\chi^\pm$  does not decay inside the detector and leaves a muon-like track but with a different mass. The existing searches for long-lived massive charged particles (CHAMPs) at LEP2 impose a bound on the mass, $m_{\chi^\pm} > 99.5$~GeV~\cite{CHAMPLEP} at 95\% C.L.. Both D0~\cite{CHAMPD0} and CDF~\cite{CHAMPCDF} at the Tevatron have also performed searches for CHAMPs. The current strongest bound is from CDF who searched for a single, isolated, weakly interacting CHAMP within the muon trigger acceptance. For the case at hand, the CDF bound constrains the sum of the production cross sections $\sigma(\chi^+\chi^-)$ and $\sigma(\chi^+\chi^0)$. The exclusion limit on the dark matter mass is $m_\chi \approx m_{\chi^\pm} \ge 121$~GeV at 95\% C.L.

For $\Delta \gtap \mathcal{O}(\mev)$ the kinetic energy of the DM is insufficient to allow an inelastic up scattering to $\chi^\pm$.  However, it is possible that for certain elements, the DM may form a short-lived electromagnetic bound state with the nucleus of an atom.  The elastic scattering of DM off a nucleus $^A_ZN$ is depicted in Fig.~\ref{fig:feynman}. The intermediate boundstate may be in its ground-state or in an excited state, depending on which state is within the accessible range of dark matter kinetic energy.  Interestingly, an $\mathcal{O}(100)\ \gev$ WIMP has kinetic energy comparable to the typical energy splitting of the energy levels of nuclei. So, if the binding energy is large enough to compensate for the splitting $\Delta$, one or two bound states may be probed in the dark matter scattering process.

For $\chi^0$ scattering off element $^{A}_ZN$ the bound state forms between $\chi^-$ and $^A_{Z+1}N$.  For large $Z$ the orbital radius is smaller than the Bohr radius of the nucleus and $\chi^-$ is bound inside the nucleus.  Using the two-parameter Fermi charge distribution~\cite{Fricke95,Duda:2006uk} for the protons inside the nucleus we calculate the binding energy of this system, by solving the Schr\"{o}dinger equation~\cite{Lucha:1998xc}.  For the case where the DM mass is much larger than the nucleus mass, the electromagnetic binding energies between $\chi^-$ and $^A_{Z+1}N$, for the relevant elements, are shown in Table~\ref{tab:bindingenergy}.  The $(\chi^-, _{Z+1}^AN)$ binding energy is small in comparison to the total binding energy of the whole nucleus and we assume that the presence of $\chi^-$ inside the nucleus does not have a large effect on the charge distribution or the spectrum of nuclear excited states.  The excited states of the bound state are $\sim 1\ \mev$ above the ground state, whereas the excited states of the nucleus alone are $\sim 500\ \kev$ for a fixed angular momentum~\cite{NuclearWeb1, NuclearWeb2}. In order to conserve angular momentum, the excited states of the nucleus alone should have angular momentum as $J_{N(Z+1, A)^*} = J_{N(Z, A)},  J_{N(Z, A)}\pm 1$ for the fermionic dark matter considered in this paper. If the excited states of the bound state can be probed, the selection rule is $|\overrightarrow{J}_{N(Z+1, A)^*} + \overrightarrow{L}+1/2 |= |\overrightarrow{J}_{N(Z, A)} +1/2|$, with $L$ as the orbital angular momentum between $\chi^-$ and $ _{Z+1}^AN^*$.
\TABLE[t!]{
\renewcommand{\arraystretch}{1.4}
\begin{tabular}{c| c c c c c c c}
\hline  \hline
$^A_ZN$ & $^{23}_{11}Na$ & $^{28}_{14}Si$ & $^{74}_{32}Ge$ & $^{127}_{\ 53}I$ & $^{129}_{\ 54}Xe$ & $^{133}_{\ 55}Cs$ &  $^{184}_{\ 74}W$  \\ \hline
$^A_{Z+1}N$ & $^{23}_{12}Mg$ & $^{28}_{15}P$ & $^{74}_{33}As$ & $^{127}_{\ 54}Xe$ & $^{129}_{\ 55}Cs$ & $^{133}_{\ 56}Ba $ & $^{184}_{\ 75}Re$  \\ \hline
$\Delta m$ (MeV) & 3.8   & 13.8  & 2.1   & 0.15 & 0.7 & 0.01 & 1.0\\ \hline
$S_n$ (MeV)        & 13.1 & 14.5  & 8.0   &  7.3   & 9.6  &  7.2   & 6.5  \\ \hline
$E_{b}$ (MeV)      &  5.8   &  7.5   & 13.5 &  19.1 & 19.4  & 19.6 & 23.3 \\
\hline  \hline
\end{tabular}
\caption{The electric binding energy for the ground state of the boundstate composed of $\chi^-$ and $^{A}_{Z+1}N$, the nuclear mass difference $\Delta m \equiv m_{^A_{Z+1}N} - m_{^{A}_ZN}$ and the neutron emission energy $S_n$. The DM mass is taken to be 500 GeV.}\label{tab:bindingenergy}}

The resonance may form in its ground state, or an excited state, and the nucleus itself may also be in an excited state.  Expressing the total excitation energy of the resonance above its ground state as $\omega$, the resonance mass is given by,
\beq
m_r\,=\,m_{\chi^-}\,+\,m_{^A_{Z+1}N}\,+\,\omega\,-\,E_b\,,
\eeq
Thus, the resonance speed squared is
\beq\label{eq:vr}
v_r^2=\frac{2(\Delta + m_{^A_{Z+1}N} - m_{^A_{Z}N} + \omega - E_b)}{\mu_{\chi N}}\,.
\eeq
The mass difference between neighboring elements, due to nuclear binding energy, depends on the elements and isotopes in question~\cite{NuclearWeb1, NuclearWeb2} and for the elements of interest lies in the range $0\ \mev \ltap m_{^A_{Z+1}N} - m_{^A_{Z}N}\ltap 14\ \mev$.  The nuclear mass differences for the corresponding atoms are listed in Table~\ref{tab:bindingenergy}.  The mass differences for nuclei are offset from the mass differences for atoms by approximately 511 keV.  The kinetic energy of the reduced mass system, $1/2\,\mu_{\chi N}\,v_r^2$ is around 200 keV.  If the splitting, $\Delta$, is comparable to the other scales in the numerator of (\ref{eq:vr}), then for some element there will be a bound state which is accessible by the DM.  Since the binding energy is dependent on $Z$, and therefore element, it is possible that some targets will have a resonantly enhanced elastic scatter and others will not, and instead have a rate  suppressed by $\delta^4/v_r^4$.  We discuss this element dependence, how it may simultaneously explain results of DAMA and the null results of other direct detection experiments, and its potential signals in more detail in Section~\ref{sec:elemental}.

Having discussed the resonance speed, we turn to discuss the various ways it can decay and estimate its decay width $\delta$.  The resonance may simply fall apart back into the initial state, namely $\chi^0$ and $^A_{Z}N$, this is the only possibility if the bound state is made in its ground state.  The width for the resonance to decay to the DM plus the original nucleus can be extracted from the cross section for $\chi^-\ ^A_{Z+1}N$ to scatter into $\chi^0\ ^A_{Z}N$,
\bea
\Gamma_{r\rightarrow \chi N}&=& \sigma v \Bigl\lvert \int f_{\chi^-}(r)\,\rho_{p}(r)\,4\,\pi\,r^2\,dr \Bigr\rvert^2  = \frac{\mu_{\chi N}\,\sqrt{2\,\Delta\,\mu_{\chi N}}}{\pi}\,\frac{G_F^2}{2}\,\Bigl\lvert \int f_{\chi^-}(r)\,\rho_{p}(r)\,4\,\pi\,r^2\,dr \Bigr\rvert^2 \nonumber  \,.
\label{eq:weakwidth}
\eea
Here, $\rho_p(r)$ is the proton charge distribution with the normalization $\int \rho_p(r) \,4\pi\,r^2 dr = Z$; $f_{\chi^-}(r)$, normalized as $\int f^2_{\chi^-}(r)\,4\pi\,r^2 dr =1$, is the wavefunction of $\chi_-$ with respect to the nucleus. For iodine, the falling-apart decay width is calculated to be 
\be
^{127}_{\ 53}I:  \hspace{2cm}\Gamma_{r\rightarrow \chi N}\approx 0.006~\mbox{eV}\,.
\ee

If the bound state forms in an excited state it has the possibility to de-excite by emission of photons and/or neutrons.  The width for emission of photons depends on the multipole moment involved in the transition.  For nuclear electric transitions of multipole moment $L$, the width is given by~\cite{nuclearBook},
\be
\Gamma_\gamma(EL) \approx \frac{8\pi(L+1)}{L[(2L+1)!!]^2}\,\left(\frac{3}{L+3}\right)^2\,\alpha\,\omega_\gamma^{2L+1}\,r_A^{2L}\,,
\label{eq:edipole}
\ee
and magnetic transitions are given by 
\bea
\Gamma_\gamma(ML)&\approx& \frac{8\pi(L+1)}{L[(2L+1)!!]^2}\, \left(\mu_p-\frac{1}{L+1} \right)^2 \frac{1}{m_p^2} \times\left(\frac{3}{L+2}\right)^2\,\alpha\,\omega_\gamma^{2L+1}\,r_A^{2L-2}~,
\label{eq:mdipole}
\eea
where $\mu_p$, typically around $3$, is the nuclear magnetic moment of the proton in nuclear magneton units.  We have assumed that the radial nuclear wavefunction has a step function profile, going to zero beyond the nuclear radius, $r_A$.  For a given $L$, magnetic transitions are negligible relative to electric transitions.  Taking the nuclear radius to be $r_A\sim 6$ fm~\cite{Duda:2006uk} and a typical transition energy of 100 keV, we find
\bea
\Gamma_\gamma(E1) \approx & \hspace{-1.2cm} 0.2\ \ev,\hspace{1cm} &\Gamma_\gamma(E2)\, \approx  3.8 \times 10^{-9}\ \ev , \nonumber \\
\Gamma_\gamma(M1) \approx  &  4.8\times 10^{-4}\ \ev,\hspace{1cm}  &\Gamma_\gamma(M2)\approx  7.4\times 10^{-11}\ \ev.
\label{eq:photonwidth}
\eea
Similarly, it is possible that rather than the nucleus changing energy level, the transitions take place due to $\chi^-$ changing energy level.  We find from solving the Schr\"{o}dinger for excited states of the bound state, with the charge distribution of the nucleus as before, that the typical radius of the $\chi^-$ orbit is 1 fm, and the typical transition energy as $\omega\sim 500\ \kev -1$ MeV.  We assume that there are no magnetic transitions, and the widths for photon emission due to electric transitions of $\chi^-$ are comparable to those of the nucleus (\ref{eq:photonwidth}).  Note that if the emitted photon energy is smaller than the kinetic energy of the incoming DM, it is possible that the excited states emits a photon and subsequently falls apart into the $\chi$ and the original nucleus.

For a heavy element the neutron and proton separation energies are comparable with $S_n\sim S_p\sim \mathcal{O}(5-10)\ \mev$.  Due to the Coulomb barrier proton emission is greatly suppressed relative to neutron emission.  If the DM-nucleus bound state forms with the nucleus in an excited state whose energy is above the neutron separation energy a neutron will be emitted from the bound state.  The width for this process is large $\Gamma_n\sim 0.1-1\ \mev$~\cite{SiemensJensen}.  The emission of the neutron may leave the nucleus in its ground state, but if not subsequent emission of photons will de-excite the nucleus.  Unlike the case of photon emission, where depending on the transition the photon energy can be less than the initial DM kinetic energy, the energy carried away by the emitted neutron is large enough that the bound state is energetically forbidden from falling apart and releasing the DM.  The difference in nuclear binding of $^A_ZN$ and $^{A-1}_ZN$ is $\mathcal{O}(\mev)$ and so is sufficiently large that the DM remains bound to nucleus $^{A-1}_ZN$ after neutron emission.

Which of the possibilities outlined above occurs depends on the details of the spectrum of nuclear excited states and DM-nucleus excited states, which are highly element dependent.  This allows for the possibility that only certain elements are capable of observing rDM whereas for others DM scattering is highly suppressed.  In particular, rDM allows the results of DAMA to be compatible with the null results from other direct detection experiments.

\section{Element Dependence and Potential Signals}
\label{sec:elemental}

One of the interesting features of the model of rDM described above is that it naturally leads to a high sensitivity to the target nucleus involved in the DM scattering.  Thus, it is possible that only for particular elements, and therefore particular experiments, there is an accessible bound state and that for others there is no rDM scattering. To make definite predictions requires detailed knowledge of the available states of the DM-nucleus bound state.  However, even without this detailed knowledge there are several general statements that can be made.
 
The experiments searching for direct detection of DM fall into two classes: those which contain only low-Z (Z$<40$) elements - CDMS (Si, Ge)~\cite{Ahmed:2008eu}, and those that contain at least one high-Z element - DAMA (Na, I)~\cite{Bernabei:2008yi}, KIMS (Cs, I)~\cite{Kim:2008zzn}, XENON (Xe)~\cite{Angle:2007uj}, CRESST (W)~\cite{Cozzini:2004vd}, and ZEPLIN (Xe)~\cite{Alner:2007ja,Lebedenko:2008gb}.  There is a large difference in binding energies between the low-Z and high-Z elements, see Table~\ref{tab:bindingenergy}, so that if the splitting $\Delta$ is large enough, $\Delta \gtap 11\ \mev$~\footnote{2.1~MeV is the mass difference of $^{33}_{74}As$ and $^{32}_{74}Ge$.}, there will be no available bound state at CDMS but there may be bound states available at the other experiments.  In this way the rate at CDMS is too small for any signal to be observed.

Presumably iodine is an element with an available bound state, since DAMA has observed a modulated signal.  If the resonance takes place in the ground state of the xenon-DM system, or if the excited state can not decay via an electric or magnetic dipole transition Eqs. (\ref{eq:edipole}-\ref{eq:mdipole}), so that the dominant width is for the bound state to fall apart into DM and the original nucleus.  Then we estimate the resonance to be very narrow, we refer to this as Case I.
\newline
\textbf{Case I:}
\bea
&&\Gamma_{tot}\approx \Gamma_{r\rightarrow \chi N} \approx 0.006~\mbox{eV}\,, \nonumber  \\
&&\delta = \sqrt{\frac{\Gamma_{tot}}{\mu_{\chi N}}} \approx 0.075~{\rm km/s}\,,  \nonumber  \\
&& \sigma_0= \frac{2J_r+1}{(2s_\chi+1)(2s_N+1)}\,\frac{4\pi^2}{\mu_{\chi N}^2} 
\frac{\delta^2}{v_r^2}\, \approx\,0.025\,~\mbox{pb}\,,
\label{eq:case1}
\eea
for $s_N=5/2$, $s_\chi=1/2$, $J_r=3$, $v_r=470$~km/s and $m_\chi = 500$~GeV. These are the numbers used to generate Fig.~\ref{fig:single}. In this case, the dark matter elastic scattering is the main process.  The inelastic process with extra gamma rays emitting is suppressed by $\Gamma_\gamma(E2)/\Gamma_{r\rightarrow \chi N}\sim 10^{-4}$.  We focus on this case as the explanation of the DAMA signal. There are only two model parameters $m_\chi$ and $\Delta$ (or $v_r$). So, we have $\chi^2/d.o.f=15.2/10=1.52$, which corresponds to a $p$-value $p=0.12$.

Alternatively, photon emission process may dominate, we refer to this as Case II.
\newline
\textbf{Case II:}
\bea
&&\Gamma_{tot}\approx \Gamma_{\gamma} \approx 0.2\times \left(\frac{\omega_\gamma}{100~\mbox{keV}}\right)^3~\mbox{eV}\,, \nonumber  \\
&&\delta = \sqrt{\frac{\Gamma_{tot}}{\mu_{\chi N}}} \approx 0.1\times \left(\frac{\omega_\gamma}{100~\mbox{keV}}\right)^{3/2}~{\rm km/s}\,,  \nonumber  \\
&& \sigma_0= \frac{2J_r+1}{(2s_\chi+1)(2s_N+1)}\,\frac{4\pi^2}{\mu_{\chi N}^2} 
\frac{\delta^2}{v_r^2}\frac{\Gamma^2_{r\rightarrow \chi N}}{\Gamma_{\gamma}^2} \, \approx\,1.4\times 10^3\times \left(\frac{\omega_\gamma}{100~\mbox{keV}}\right)^{-3}\,~\mbox{pb}\,.
\label{eq:case2}
\eea
for $s_N=5/2$, $s_\chi=1/2$, $J_r=3$. The DAMA modulated spectrum can also be fitted by choosing $\omega_\gamma=50$~keV, $v_r=450$~km/s and $m_\chi = 250$~GeV ($\chi^2=15.6$ and $p=0.11$). If the photon emission width does not completely dominate the width to fall apart, this situation may still be capable of explaining the DAMA excess, and in addition there will be correlated signals which can be searched for.   Since the elastic dark matter scattering is not the main process, Case II predicts photons with a few hundred keV energy at DAMA.  These gamma rays should have energy matched to a nuclear spectral line of the element $^{127}_{\ 54}Xe$. The number of gamma rays events will be larger than the number of modulated events by a factor of $\Gamma_\gamma/\Gamma_{r\rightarrow \chi N}\sim 10-100$.  If the energy of the emitted photon is always below the kinetic energy of the incoming DM the bound state can emit a photon and then fall apart.  Again, along with the DAMA signal we would expect photon lines.  But if the emitted photon energy is larger than the DM kinetic energy then the DM will remain permanently bound to the nucleus, and it is not possible to explain the DAMA excess.

The binding energy grows with Z and it is possible that for some high-Z element the difference $E_b-\Delta$ is large enough that the resonance, if it occurs, is above the neutron separation energy $S_n\sim 5-10\ \mev$; the process of proton emission from the nucleus requires tunneling through the Coulomb barrier and is suppressed relative to neutron emission, we do not consider the case of proton emission here.  For such an element a dark matter scatter has very different kinematics from the usual elastic scatter.  The recoil energy is determined not only by the kinetic energy of the incoming dark matter $\sim 100\ \kev$ but receives contributions from the energy released when the neutron is emitted, $E_b-\Delta-S_n\sim \mev$.  Such a recoil is outside of the usual energy range searched for at conventional direct detection experiments and may require dedicated experiments to find.  If the resonance level is above the neutron separation energy this decay channel dominates and the DM remains bound to the nucleus.  

While it is possible, for $11\ltap \Delta/\mev\ltap 16$, that there will be no events at CDMS\footnote{Although there is no tree-level elastic scattering there can be scattering at the loop level.  The rate for this is highly suppressed, $\sigma_p\sim 10^{-45}$ cm$^2$,~\cite{Cirelli:2005uq} and will not be observable until the next generation of DM direct detection experiments.} and the scattering off tungsten will involve neutron emission and likely be unobservable at CRESST.  The proximity of binding energies for experiments involving iodine, caesium and xenon present more of a challenge.  KIMS in particular contains the same element as DAMA and there is no confusion as to the position of the resonant velocity.  As a result the KIMS results place strong constraints on rDM, see Fig.~\ref{fig:allowedregions}.

The xenon based experiments of ZEPLIN and XENON may also present strong constraints.  The splitting of the appropriate energy levels in $^{129}_{\ 55}Cs$ is $\sim 500\ \kev$, so it is possible that no resonance is available for these xenon experiments.  
However, it is difficult to be sure due to the comparable binding energies of iodine and xenon and the preponderance of different isotopes of xenon present in the detectors.  For simplicity we have concentrated throughout on one of the high abundance isotopes.  Furthermore, rDM has the feature of increased modulation (see Fig.~\ref{fig:modulationratio}) and the XENON data was taken between October and February, which somewhat weakens their bounds on rDM.  Similar to the case II for DAMA, the spectral lines corresponding to the isotopes of cesium may be the dominant signals at the ZEPLIN and XENON. A more detailed study of the spectra and number of the predicted events in other experiments and the allowed resonance speeds for each experiment is warranted.

It is amusing to consider the possibility that the signal in DAMA is not coming from scattering off iodine but instead from scattering off sodium.  For $\Delta\ltap 2\ \mev$ it is possible that in all other experiments, with the exception of silicon for which there is no resonance, that there is neutron emission.  We have been unable to find parameters within our simple model that allow this to work whilst keeping the dark matter mass above the collider bounds mentioned earlier, but it is an intriguing possibility within the rDM framework.

\subsection{Other Constraints and Signals}

The process of DM capture with neutron, or photon, emission 
may occur for many high-Z elements, and so searches for anomalously heavy elements place a constraint on models of this type.  However, very few of the searches for anomalously heavy elements have been with high-Z elements, there have been some searches for heavy nuclei of gold and iron using mass spectroscopy~\cite{Javorsek:2002bf} with bounds on the allowed fraction of $f_{Au} \ltap \mathcal{O}(10^{-10})$ and $f_{Fe} \ltap \mathcal{O}(10^{-8})$, for a 500 GeV DM particle.  Assuming that an appropriate energy level exists the expected fraction of element $X$ that will have captured a DM particle during exposure time $\tau$, taken to be of order the age of the Earth, is
\be
f_X= \langle \sigma v \rangle n_\chi \tau\approx 6\times 10^{-13} \left(\frac{\langle \sigma v \rangle}{3\times 10^{-26}\,\mathrm{cm}^3\,\mathrm{s}^{-1}}\right) \left(\frac{\tau}{10^9\,\mathrm{yrs}}\right) \left(\frac{500\ \gev}{m_\chi}\right)~,
\ee 
where for the ambient dark matter density we take $n_\chi=0.3\ \gev/m_\chi$ cm$^{-3}$.  Using the delta-function limit of (\ref{eq:xsecformula}) we find that the effective cross section is given by,
\be
\langle \sigma v\rangle =\frac{1}{2\,\sqrt{\pi}}\frac{\sigma_0\,v_r}{v_0 v_E}\left( e^{-(v_r-v_E)^2/v_0^2} - e^{-(v_r+v_E)^2/v_0^2} \right)
~.
\ee 
Over the whole parameter space shown in Fig.~\ref{fig:allowedregions} the fraction of heavy elements that have captured DM is just below the present bound. For example, 
the Case I in Eq.~(\ref{eq:case1}) has $f_X\approx 4\times 10^{-12}$. However, an improvement in the bound would start to probe much of the parameter space.  Moreover, an extension of these searches to other high-Z elements would shed light on the splitting $\Delta$ and which elements have accessible energy levels.

In the simple model described in Section~\ref{sec:model} the DM is part of a weak triplet.  As mentioned earlier the present collider bounds, from the Tevatron, on the charged state in the triplet puts the mass scale above $\sim 120\ \gev$.  For masses close to this bound, $m_\chi=150\ \gev$, the cross section for pair production of the charged states at the LHC is large.  At $s^{1/2}=10\ \tev$ it is 180 fb and at $s^{1/2}=14\ \tev$ it is 280 fb, but it drops to $\sim 1$ fb for $m_\chi=500\ \gev$ at $s^{1/2}=14\ \tev$.  It can be searched for as long-lived charged states at the LHC.

The main dark matter annihilation products are $W^\pm$ gauge bosons, and the typical annihilation cross section is order of 1 pb. However, if the splitting, $\Delta$, is small enough that there is an available $(\chi^+\chi^-)$ resonance the DM  annihilation cross section may be greatly enhanced~\cite{Hisano:2006nn,Cirelli:2007xd,Pospelov:2008qx}.  
In this case, lots of positrons, anti-protons, photons and neutrinos can be produced and contribute to cosmic rays.

\section{Discussions and Conclusion}
\label{sec:conclusion}

Resonant DM provides a new scenario to explain the DAMA modulation results, while being consistent with other dark matter direct detection constraints. The model presented in this paper is very simple: a dark matter particle, which is part of an electroweak triplet which has an ${\cal O}(10)$~MeV mass splitting between its neutral and charged components, can realize all the features necessary for rDM. In our model the resonance present in the DM-nucleus scattering is an electromagnetic bound state of the charged partner of the DM and a nucleus.  The simple model presented here may not be unique, other models can also be constructed to realize rDM, if there contain a bound state of dark matter and the nucleus. 

There are two novel features of rDM. Firstly, the resonance effect picks out from the Maxwell-Boltzmann dark matter velocity distribution a narrow window of DM velocities around the specific resonance velocity. If this velocity happens to be in the high-speed tail of the velocity distribution, the modulation amplitude of DM scattering can be enhanced. Secondly, rDM depends on many detailed properties of nuclei. Other than the mass, $Z$ and $A$ of a nucleus, the resonance speed can also depend on the energies of its various excitation levels.  It is conceivable that the only element capable of direct detection of the DM in the halo is iodine.  Because of this the KIMS experiment is in an ideal situation to confirm or deny the DAMA excess and the rDM model.  

The elemental dependence of the scattering rate in rDM means that it is not straightforward to make precise predictions for the event rate at other experiments, it would be worthwhile to address the allowed parameter space for the other experiments in a model independent fashion.  However, it is straightforward, to accommodate the null results at CDMS.  The model predictions for ZEPLIN and XENON depend highly on a more precise calculation of the binding energy and more accurate values for nuclei energy levels. For CRESST, depending on the mass difference of the DM particle and its partner, one can either have lots of MeV neutron events or a suppressed number of nuclear recoil events if there is no rDM effect.  Exactly what occurs at DAMA and KIMS depends on which state the rDM effects probe.  We have two cases:  case one has only elastic scattering, if the electric and magnetic dipole radiation is forbidden, for instance if the bound state forms in its ground state.  While case two predicts lots of gamma rays in addition to the elastic scattering events. The energy of those gamma rays should correspond to a nuclear transition in $^{127}_{\ 54}Xe$.  It would be interesting to search for this signal at existing direct detection experiments.

We have surveyed the allowed parameter space of rDM in a model independent fashion and have found it compatible with the DAMA and KIMS data for a wide range of DM mass, $m_\chi\gtap 100\ \gev$.  We have focused on the case of one resonance but note that there can in principle be several resonances available.
For our concrete model, where the bound state requires a  charged state nearly degenerate with the DM, there are colliders bounds on the charged state requiring $m_\chi\gtap 120\ \gev$.  This charged partner can be searched for at the Tevatron and the LHC.  In addition, the DM can form stable bound states with heavy elements, the capture rate is just beyond the present bound.  However, this search has  only been carried out for a few heavy elements and by extending these searches for anomalously heavy nuclei to a larger set of heavy elements one can place strong constraints on rDM and probe the element dependence.  In rDM models, the search for heavy nuclei is highly correlated with DM direct detection searches.  Discovery of an anomalously heavy element would identify what material should be used for future direct detection experiments.

In conclusion, the resonance effects can dramatically change the ``traditional"  dark matter \emph{elastic} direct detection calculation. Resonant DM can enhance the modulation effect and relies strongly on the detailed nuclear properties of different elements. The modulated data at DAMA can be explained and the constraints from unmodulated data can be satisfied. Furthermore, the rDM can reconcile the apparent contradiction between DAMA and other experiments like CDMS, XENON, KIMS, ZEPLIN and CRESST. We have described a simple model, which can realize all the features of rDM.  It has dark matter, with $m_\chi\gtap 120\ \gev$, to be part of an electroweak triplet which has a mass splitting between its neutral and charged parts of around 10~MeV.  

\acknowledgments 
\vspace*{-.1in}
Many thanks to Spencer Chang, Scott Dodelson, S. K. Kim, Kaixuan Ni for interesting discussions.  We thank Maxim Pospelov for stimulating discussions and reading a draft of this paper.  Fermilab is operated by Fermi Research Alliance, LLC under contract no. DE-AC02-07CH11359 with the
United States Department of Energy. 

\bibliography{RDMbib}
\bibliographystyle{JHEP}

\end{document}